\documentclass[12pt,twoside]{article}
\usepackage{xrb2007}
\usepackage{psfig}
\usepackage{epsf}
\usepackage{graphics}
\usepackage{graphicx}
\usepackage{lscape}
\begin{document}
\def\4u{4U\,1822-371}
\session{Faint XRBs and Galactic LMXBs}
\shortauthor{Bayless et al.}
\shorttitle{ADC Source 4U\,1822-371}
\title{Optical and UV Light Curves of the Accretion Disk Corona
Source 4U 1822-371} 
\author{Amanda J. Bayless, Edward L. Robinson, and Mark E. Cornell}
\affil{Department of Astronomy, University of Texas at Austin
1 University Station, Austin, TX 78712-0259}
\author{Robert I. Hynes and Teresa A. Ashcraft}  
\affil{Department of Physics and Astronomy, Louisiana State University,Baton Rouge, LA 70803}
\begin{abstract}
The eclipsing low-mass X-ray binary \4u is the prototypical 
accretion disk corona (ADC) system.  
We have obtained new time-resolved UV spectrograms of \4u with
the Hubble Space Telescope and new $V$-
and $J$-band light curves with the 1.3-m SMARTS telescope at CTIO.  
We present an updated ephemeris for the times of the optical/UV eclipses.
Model light curves do not give acceptable fits to the UV eclipses
unless the models include an optically-thick ADC.
\end{abstract}

\section{Introduction}
An accretion disk corona in a low-mass X-ray binary (LXMB)
is material above and below the accretion disk that emits
hard X-rays.
It is believed that there is an ADC in most, if not all, LMXBs, but the ADC
is invisible in most systems because the neutron star and
inner disk are much more luminous.  The ADC is visible in LMXBs with nearly edge-on
orbital
inclinations so that the X-ray flux from the neutron star and
central disk are blocked by other parts of the disk further out from the 
center \citep{whi82}. 
Because systems in which the ADC is visible have high inclinations, 
the accretion disk and ADC are eclipsed by the secondary star.
The eclipse light curve yields information about
the geometry and spectral energy distribution of the ADC.

\4u is the prototypical ADC system \citep{mas80}.  The existence of
the ADC in \4u was deduced from the partial X-ray eclipse, 
which has two
notable properties: (1) it has a gradual ingress and egress and is 
relatively shallow, with 50\% of the flux
remaining at mid-eclipse, and (2) it is wide, typically
lasting $\Delta \phi = 0.1$ in orbital phase at X-ray wavelengths
and double that at optical wavelengths \citep{hellier90}.
\citet{whi81} showed that the X-ray emitting region must be
vertically extended and comparable in size to the secondary star
to produce the observed X-ray eclipse. 
The ADC is optically thick at X-ray wavelengths.

\section{Observations}
\begin{figure}
\center
\includegraphics[scale=0.35, angle=270]{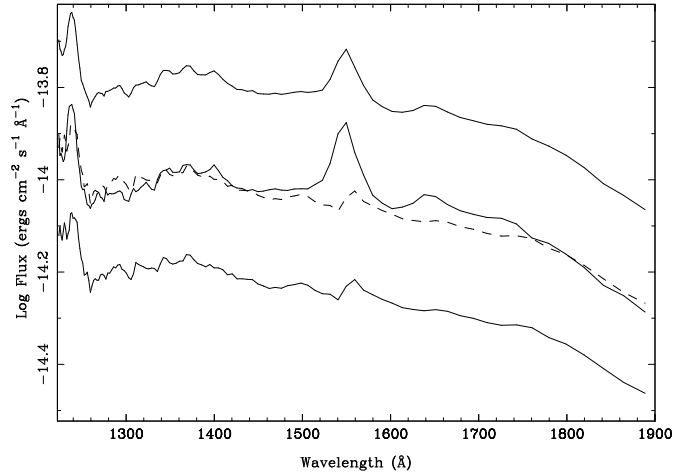}
\caption{The UV spectrum of \4u.  The top line is the
  mean spectrum, the middle solid line is the spectrum for orbital
  phases within $\pm0.05$ of mid-eclipse, the bottom line is the eclipse 
  spectrum subtracted
  from the average spectrum, and the middle dashed line is the subtracted
  spectrum shifted up to the eclipse spectrum.  
  The N~{\small V} emission line at 1240~\AA\ is mostly eclipsed, while 
  the C~{\small IV}~1550~\AA\ and He ~{\small II}~1640~\AA\ line are 
  mostly not eclipsed.}
\label{fig1}
\end{figure}

The new observations of \4u consist of time-series UV spectrophotometry 
obtained in 2006 with the Hubble Space Telescope using an
objective prism on the ACS/SBC \citep{acs}.  
The spectrograms have a usable wavelength range from 1222~\AA\ to 1900~\AA.
The mean spectrum, shown in Figure~\ref{fig1}, has prominent emission lines
of N~{\small V} at 1240~\AA, a blend of O~{\small IV}, O~{\small V}, 
and Si~{\small IV} near 1370 \AA, C~{\small IV} at 1550~\AA, and 
He~{\small II} at 1640~\AA.  
There are also ISM absorption blends of Si~{\small II} and C~{\small I}
near 1260~\AA, and of O~{\small I} and Si~{\small III} near 1300~\AA.  
We formed an eclipse spectrum by averaging all
spectrograms near mid-eclipse, and then a spectrum of the eclipsed 
flux by subtracting the eclipse spectrum from the mean spectrum (Figure \ref{fig1}).
The N~{\small V} line is mostly eclipsed while the C~{\small IV} 
and He~{\small II} lines are not.  
The C~{\small IV} and He~{\small II} emission must come from far above
and below the disk.

The optical observations of 4U 1822-371 were taken in the $V$- and $J$-bands
with ANDICAM on the SMARTS 1.3-m telescope at CTIO \citep{andicam}.  
The observations consisted one or two measurements per night 
during 30 June - 15 September 2005 and 
29 March - 20 October 2006.

\section{Data Analysis}
\begin{figure}
\plottwo{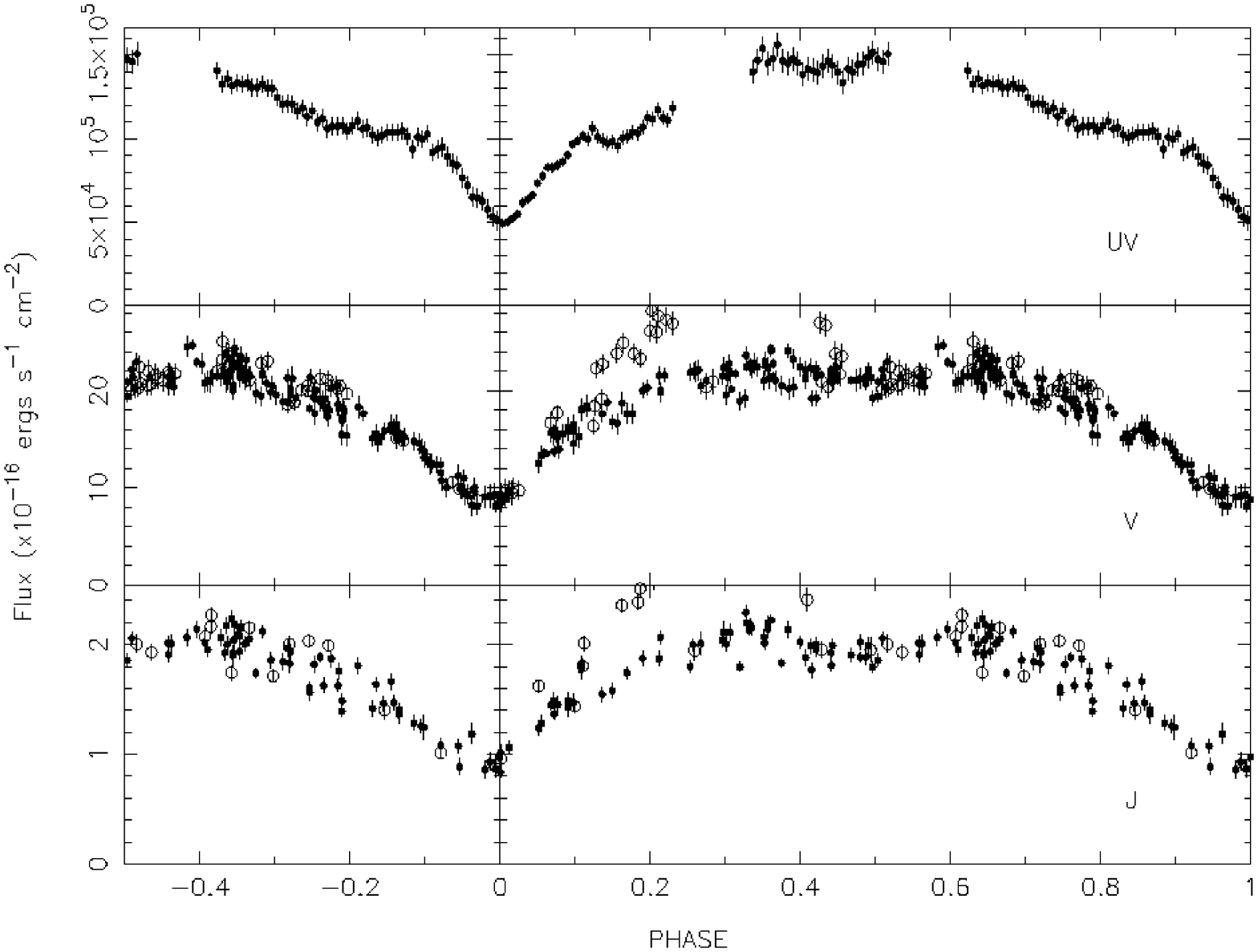}{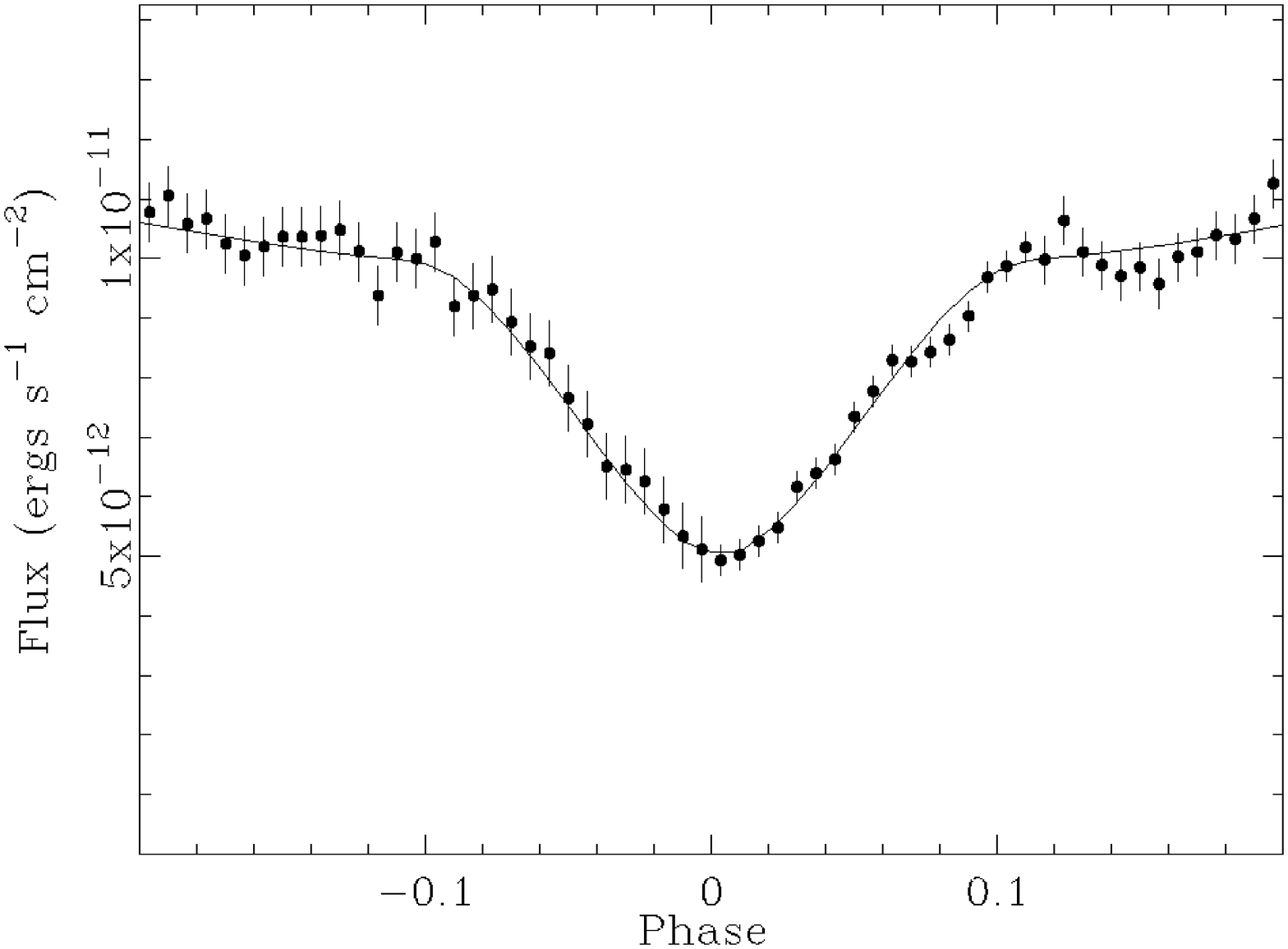}
\caption{{\it Left Panel:} The mean light curve for the integrated UV flux, 
  and the folded $V$- and $J$-band light curves.  The open circles are
  from 2005 and the solid circles from 2006.  {\it Right Panel:} The UV
  eclipse light curve. The points
  are the observed light curve and the solid
  line is the light curve produced by a model with an optically thick 
  toroidal ADC.}
\label{fig2}
\end{figure}

The UV spectrophotometry yielded two complete eclipse light curves.
Because of the sparse sampling, the $V$- and $J$-band data
were folded at the orbital period to produce mean light curves 
for the two observing seasons.  
Joining the new eclipse times to published times for the
optical/UV eclipses, we find the following ephemeris for
the times of minimum of the optical/UV eclipses
\begin{displaymath}
T_{min} = \textrm{HJD}\ 2445615.31166(74) 
           + 0.232108641(80)E+2.46(21)\times10^{-11}E^2,
\end{displaymath}
where E is the cycle number.  
The left panel of Figure \ref{fig2} shows the 
UV, $V$, and $J$ light curves phase folded on this ephemeris.

We modeled the eclipse light curve from orbital phase -0.2 to 0.2,
constraining the inclinations to $i>80^{\circ}$ 
because the neutron star is hidden from view. 
We attempted to fit the eclipse with a wide variety of disk models without
success.  These models included flat disks with uniform surface brightness,
flared disks, flared disks with a vertically-extended rims, 
alpha-model disks, and all of these with irradiation. 
A few models failed because they gave light curves with clearly
incorrect
shapes, but most of the models failed in one of two generic ways: 
the eclipse light curves were too deep because the center of the disk
was producing too much flux, or they required an inclination less than 
$80^\circ$.
To remedy these defects we added a quasi-toroidal ADC.  
The ADC must be optically thick and vertically extended to block flux 
from the neutron star and inner disk.
The right panel of Figure \ref{fig2} shows the best fitting
model light curve.  
In this model the orbital inclination is $i=83.5^{\circ}$ and
the ADC extends out to $\sim 1/2$ the disk 
radius and has a maximum height also $\sim 1/2$ the disk radius.

\section{Conclusion}
To obtain acceptable fits to the UV eclipse light curve of
\4u it was necessary to include an optically thick ADC 
in the light curve models. 
The ADC is large, with a
radius and height equal $\sim 1/2$ the disk radius.  
Most of the light blocked during mid-eclipse in these
models comes from
the ADC, so the subtracted spectrum shown in Figure~\ref{fig1}
is the UV spectrum of the ADC.
The spectrum is not greatly different from the average spectrum of
the system exception that
the  C~{\small IV} and He~{\small II} emission lines are nearly
uneclipsed.


\begin{thebibliography}{}
\expandafter\ifx\csname natexlab\endcsname\relax\def\natexlab#1{#1}\fi
\bibitem[{{DePoy} {et~al.}(2003)}]{andicam}
{{DePoy}, D.~L., et al.} 2003, Proceedings of the SPIE, 4841, 827
\bibitem[Hellier et al.(1990)]{hellier90}
Hellier, C., Mason, K. O., Smale, A. P., \& Kilkenny, D. 1990,
 \mnras, 224, 39P
\bibitem[{{Mason} {et~al.}(1980){Mason}, {Seitzer}, {Tuohy}, {Hunt},
  {Middleditch}, {Nelson}, \& {White}}]{mas80}
{Mason}, K.~O., {Seitzer}, P., {Tuohy}, I.~R., {Hunt}, L.~K., {Middleditch},
  J., {Nelson}, J.~E., \& {White}, N.~E. 1980, \apjl, 242, L109
\bibitem[{{Pavlovsky} {et~al.}(2006)}]{acs}
{{Pavlovsky}, C., et al.} 2006, {Advanced Camera for Surveys Instrument
  Handbook for Cycle 16, Version 7.1} (Baltimore: STScI)
\bibitem[{{White} {et~al.}(1981){White}, {Becker}, {Boldt}, {Holt},
  {Serlemitsos}, \& {Swank}}]{whi81}
{White}, N.~E., {Becker}, R.~H., {Boldt}, E.~A., {Holt}, S.~S., {Serlemitsos},
  P.~J., \& {Swank}, J.~H. 1981, \apj, 247, 994
\bibitem[{{White} \& {Holt}(1982)}]{whi82}
{White}, N.~E. \& {Holt}, S.~S. 1982, \apj, 257, 318
\end{thebibliography}
\end{document}